\documentclass[preprint,showpacs,preprintnumbers,amsmath,amssymb,pre]{revtex4}
\usepackage[english]{babel}
\usepackage[dvipdf]{graphicx}
\usepackage[dvips]{epsfig}
\usepackage{amssymb}
\usepackage{amsmath}
\usepackage{amsbsy}
\usepackage{dcolumn}% Align table columns on decimal point

\usepackage[section]{placeins}
\usepackage{color}
\newcommand {\um} {~\mu \mathrm{m}}
\newcommand {\Cmin} {~{^\circ}\mathrm{C}~\mathrm{min}^{-1}}

\begin{document}

\title{Nucleation and growth of micellar polycrystals under time-dependent volume fraction conditions}

\author{Ameur Louhichi$^{1,2}$, Elisa Tamborini$^{1,2}$, Neda Ghofraniha$^{1,2}$, Fran\c{c}ois Caton$^{3}$, Denis Roux$^{3}$, Julian Oberdisse$^{1,2}$, Luca Cipelletti$^{1,2}$ and Laurence Ramos$^{1,2,\star}$ }

\affiliation{
$^1$Universit\'{e} Montpellier 2, Laboratoire Charles Coulomb UMR 5221, F-34095, Montpellier,
France\\
$^2$CNRS, Laboratoire Charles Coulomb UMR 5221, F-34095, Montpellier, France\\
$^3$ Laboratoire de Rh\'eologie et Proc\'ed\'es UMR 5520, Universit\'e Joseph Fourier, Grenoble INP, CNRS,
Domaine Universitaire BP53,
F-38041 Grenoble Cedex 09, France}

{\small* Email: laurence.ramos@univ-montp2.fr}
\date{\today}

\begin{abstract}

We study the freezing kinetics of colloidal polycrystals made of micelles of Pluronic F108, a thermosensitive copolymer, to which a small amount of silica nanoparticles of size comparable to that of the micelles are added. We use rheology and calorimetry to measure $T_c$, the crystallization temperature, and find that $T_c$ increases with the heating rate $\dot{T}$ used to crystallize the sample. To rationalize our results, we first use viscosity measurements to establish a linear mapping between temperature $T$ and the effective volume fraction, $\varphi$, of the micelles, treated as hard spheres. Next, we reproduce the experimental $\dot{T}$ dependence of the crystallization temperature with numerical calculations based on standard models for the nucleation and growth of hard spheres crystals, classical nucleation theory and the Johnson-Mehl-Avrami-Kolmogorov theory. The models have been adapted to account for the peculiarities of our experiments: the presence of nanoparticles that are expelled in the grain boundaries, and the steady increase of $T$ and hence $\varphi$ during the experiment. We moreover show that the polycrystal grain size obtained from the calculations is in good agreement with light microscopy data. Finally, we find that the $\varphi$ dependence of the nucleation rate for the micellar polycrystal is in remarkable quantitative agreement with that found in previous experiments on colloidal hard spheres. These results suggests that deep analogies exist between hard-sphere colloidal crystals and Pluronics micellar crystals, in spite of the difference in particle softness. More generally, our results demonstrate that crystallization processes can be quantitatively probed using standard rheometry.

\end{abstract}
\maketitle

\section{Introduction}

The crystallization of colloidal suspensions has been investigated in detail because of its intrinsic interest (e.g. in protein crystallization~\cite{piazza00}), and as a means to understand crystallization in atomic and molecular systems, for which colloids are often regarded as a model system~\cite{pusey86} whose characteristic time and length scales are more readily accessible. Thanks to advanced light scattering techniques and scanning confocal microscopy, experiments have provided unprecedented information on the nucleation and growth of crystals, including the volume fraction dependent nucleation rate or the structure of the nuclei \cite{Schatzel93,Harland97,Henderson98,Palberg99,Sinn01,Gasser01,Cheng01,Martin03,Herlach10}. Most experiments have focussed on hard-sphere or charged colloids, although more recently more complex systems, such as thermosensitive colloidal microgels or mixtures of colloids and polymers have emerged \cite{Okubo02,Meng07,Palberg09,Savage09,Muluneh12}. In most experiments on colloids, crystallization has been investigated following a quench from the fluid phase to the crystal phase. This is usually achieved by shear melting the samples and then stopping abruptly the shear \cite{Henderson98,Gasser01,Schope06,Engelbrecht11,Muluneh12}, or by changing abruptly the temperature for mixtures of colloids and thermosensitive surfactant micelles~\cite{Savage09} that induce attractive depletion interactions. To the best of our knowledge, a slow transition from the fluid to the crystalline phase has not been investigated yet, in spite of the obvious relevance of such a protocol for the comparison with atomic and molecular materials. In fact, such an experiment would require tuning precisely and \textit{in situ} the particle volume fraction, which is in general a very challenging task. Thermoswellable microgel particles could in principle be used~\cite{SessomsPTRSA2009}, but typical swelling ratios are too small to allow a wide range of volume fractions to be covered by a sample prepared at a single microgel concentration. A different approach has been proposed in Ref. \cite{Merlin12}, where the slow evaporation of the solvent in a microfluidic chip is exploited. In this paper, we use the temperature-dependent formation of spherical micelles in Pluronics block copolymer solutions to investigate the crystallization kinetics in a colloidal system whose volume fraction is increased at a constant rate.

Triblock copolymers of Pluronics type are a class of commercial polymers made of two lateral water-soluble polyethylene oxide blocks and a central polypropylene oxide block, whose degree of hydrophobicity can be tuned by varying the temperature, $T$. Depending on polymer concentration and temperature, the Pluronics copolymer can form spherical micelles that are sufficiently monodisperse to crystallize at high volume fraction. Due to their amphiphilic character, thermoresponsive properties, and biocompatibility, the Pluronics copolymer are extensively used in industrial applications as antifoaming or thickener agents for instance,  and have been recognized as having potentials in biomedical sciences \cite{Xiong06,Batrakova08}. Because these materials exhibit the viscoelastic properties of gels at room temperature and can flow at lower temperature, they have also been considered as potentially interesting hydrogels in electrophoresis experiments \cite{Wu97,Rill98,Svingen02,Svingen04}. Pluronics samples are also at the heart of numerous more fundamental investigations. Depending on the lengths of the different building blocks, the temperature, and the concentration, Pluronic copolymer can self-assemble and form liquid crystalline and crystalline phases in aqueous solvents \cite{Alexandridis99,Mortensen01}. In particular, crystalline phases made of spherical polymer micelles arranged on cubic lattices in  water are frequently obtained. These phases have been used to template a crystalline assembly of nanoparticles \cite{Pozzo05,Pozzo07} or as convenient systems to investigate the non-linear rheology and flow properties of polycrystals \cite{Molino98,Eiser00}. However, their crystallization kinetics have been hardly investigated.

We have recently taken advantage of the thermophysical properties of Pluronics micelles to study the effect of the thermal history of Pluronics micellar crystal on the microstructure of polycrystalline samples. We have shown using confocal microscopy \cite{Ghofraniha12} that the average grain size of Pluronics polycrystals can be tuned by changing the heating rate, a property well established for molecular and atomic systems, but hardly investigated in colloidal materials. In this paper, we deepen the experimental study of the role of the heating rate on Pluronics concentrated samples \cite{Meznarich11}. We use standard rheometry to follow the crystallization processes while $T$ is continuously increased at a fixed rate, showing that varying the temperature of the Pluronics sample results in a continuous increase of the volume fraction of the micelles. The crystallization temperature is found to increases with $\dot{T}$: this behavior is rationalized using standard theories for the nucleation and growth of crystals developed for atomic and molecular systems and extended to colloidal materials. Our analysis allows kinetics crystallization parameters to be accessed and compared to experimental data for hard-sphere suspensions. Finally, we provide a quantitative link to the microstructure of the colloidal polycrystal, a property hardly explored in colloidal crystals~\cite{Auer05}.

The paper is organized as follows. We describe the system and the experimental techniques in Sec.~\ref{Sec:MM}, and present our results in Sec.~\ref{Sec:Results}. Section~\ref{Sec:Model} is devoted to a detailed presentation of the modeling of the nucleation and growth of crystallites under non-isothermal conditions and of the resulting polycrystal texture. Finally, in Sec.~\ref{Sec:Analysis} we quantitatively analyze our experimental results in the framework of our model and critically compare them to numerical and experimental results for colloidal hard-sphere suspensions.

\section{Materials and Methods}
\label{Sec:MM}

\subsection{Samples}

The copolymer polycrystals are composed of an aqueous suspension of Pluronic F108, a commercial PEO-PPO-PEO triblock copolymer purchased by Serva Electrophoresis GmbH, where PEO and PPO denote polyethylene oxide and polypropylene oxide, respectively. Each PEO block is made of $132$ monomers, and the central PPO block is made of $52$ monomers. The copolymer mass fraction is fixed at $34\%$. The copolymer is fully dissolved at $T \simeq 0{^\circ}\mathrm{C}$; upon heating, the PPO central block becomes increasingly hydrophobic, resulting in the formation of micelles with a diameter of $22$ nm \cite{Alexandridis95}, whose number increases with $T$, eventually leading to a crystalline phase at room temperature due to the packing of the micelles \cite{MortensenPRL92}. The copolymer polycrystals are seeded with $1$ $\%$ volume fraction nanoparticles. We used Bindzil$^\circledR$ plain silica nanoparticles (kind gift from Eka Chemical, sample type 40/130), with an average diameter of $30$ nm and a relative polydispersity of $19\%$, as determined by transmission electron microcopy. We have recently shown by small angle neutron scattering~\cite{Tamborini12} that the presence of nanoparticle (NPs) does not perturb the crystalline order of the copolymer micelles.

\subsection{Experimental techniques}

Rheology measurements have been performed using a stress-controlled rheometer (Physica UDS 200) equipped with a Couette cell. A thin layer of low viscosity silicon oil is spread on top of the sample to prevent water evaporation. Temperature is controlled by a circulating water bath. A temperature ramp from $3$ to $23{^\circ}\mathrm{C}$ at a given rate, $\dot{T}$, is imposed to the circulating water. The actual sample temperature has been checked for each imposed $\dot{T}$ by inserting a temperature probe directly in the sample confined between the cup and the bob of the Couette cell. Typically, $\dot{T}$ has been varied between $10^{-3}\Cmin$ and $2\Cmin$. For the fastest ramps, $\dot T \geq 0.007\Cmin$, the same rate is applied in the full range of temperatures, $3~{^\circ}\mathrm{C} < T < 23~{^\circ}\mathrm{C}$. In order to reduce the time required to prepare a sample, slower ramps are run in three steps: a relatively fast ramp $\dot T=0.5\Cmin$ is first imposed, up to $T=12~{^\circ}\mathrm{C}$. The desired slow ramp at $\dot T \le 7\times 10^{-3}\Cmin$ is then imposed in the intermediate temperature range $12~{^\circ}\mathrm{C}-17~{^\circ}\mathrm{C}$, during which the sample is fully solidified. Finally, the temperature is raised to $23{^\circ}\mathrm{C}$ at a rate $\dot T= 7\times 10^{-3}\Cmin$. We have checked that such a composite ramp provides the same rheological response and crystallization temperature as a non-composite ramp with a fixed temperature change rate, equal to that imposed in the intermediate temperature range from $12$ to $17{^\circ}\mathrm{C}~\mathrm{min}^{-1}$.

Complementary calorimetric experiments have been performed on a
Micro-DSC III from Setaram, using deionized water as a
reference sample. Cooling and heating ramps between $0.5$ and
$25.7~{^\circ}\mathrm{C}$ at different imposed speeds, $\dot{T}$,
have been used. An isothermal step ($1$ h at
$0.5{^\circ}\mathrm{C}$) is used to thermally equilibrate the heat
flow before starting the heating ramp, and an isothermal step ($900$
s at $25.7{^\circ}\mathrm{C}$) is used before starting the cooling
ramp. In the following, we show the data acquired during the heating
ramps.

Sample imaging has been performed with an upright Leica microscope
equipped with an air x63 objective using differential interference
contrast. The samples are confined in chambers made by two
coverslips separated by a $250  \um$m thick $16 \times 16
~\rm{mm}^2$ double-adhesive gene frame (Thermo Scientific). Samples
are prepared by placing the sample chamber in a copper container
immersed in a Haake thermal bath, whose temperature is raised from
$3~{^\circ}\mathrm{C}$ to $20~{^\circ}\mathrm{C}$ at a controlled
temperature rate.

%-----------------------------------  FIG1 -------------------------------------
\begin{figure}
\includegraphics[width=0.45\textwidth]{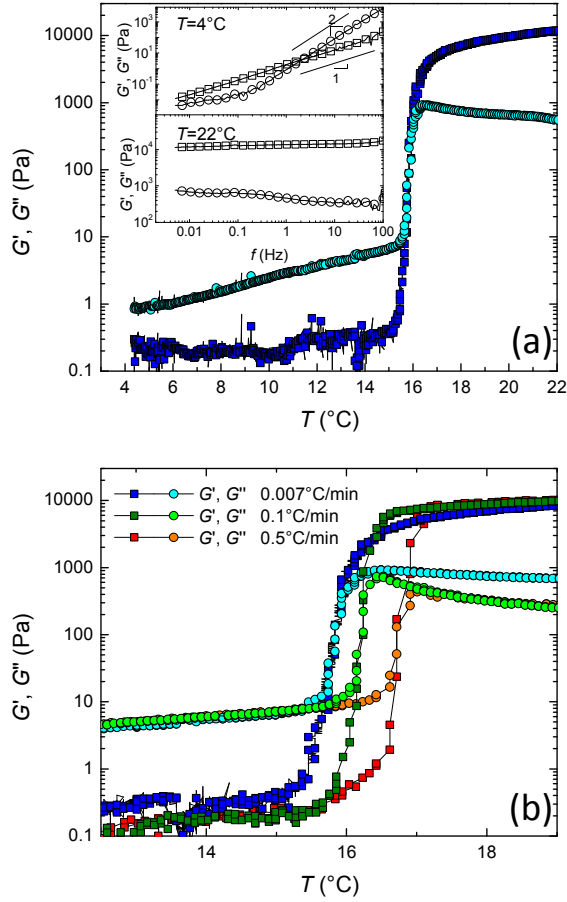}
\caption{(Color online) Storage (squares) and loss (circles) moduli as a function of temperature during a temperature ramp at a fixed positive rate. In (a) the temperature rate is $0.007\Cmin$, in (b) data are shown for three different temperature rates ($0.007$, $0.1$ and $0.5\Cmin$). The strain amplitude is $0.1 \%$ and the frequency $0.5$ Hz. Inset: frequency dependence of the storage (squares) and loss (circles) moduli in the fluid phase at low temperature (top) and in the crystalline phase at room temperature (bottom).}
\label{fig:1}
\end{figure}
%------------------------------------ FIG1 --------------------------------------

\section{Experimental Results}
\label{Sec:Results}

\subsection{Mapping between temperature and volume fraction}

The complex modulus in the linear regime is measured as a function
of time at a fixed frequency, $f=0.5$ Hz, while a temperature ramp
is imposed at rate $\dot{T}$, from  $T=3~{^\circ}\mathrm{C}$ to
$T=23~{^\circ}\mathrm{C}$ (fig.~\ref{fig:1}a). At low temperature
the sample is fluid-like with a loss modulus, $G''$, of the order of
$1$ Pa and a storage modulus, $G'$, of the order of $0.1$ Pa. Above
$\sim 4~{^\circ}\mathrm{C}$, $G''$ increases smoothly while $G'$ is
constant. This regime is followed by an abrupt increase of both $G'$
and $G''$, and the storage modulus eventually exceeds the loss
modulus. At the final temperature $G'$ is of the order of $12000$ Pa
and is at least one order of magnitude larger than $G''$. The very
abrupt fluid-to-solid transition is the signature of the sample
crystallization. We define $T_c$ as the temperature at which the
storage and the loss modulus cross each other, and identify $T_c$ as
the crystallization temperature. We show in the inset of
fig.~\ref{fig:1}a the frequency dependence of the complex modulus at
low ($T=4{^\circ}\mathrm{C}$) and room ($T=23{^\circ}\mathrm{C}$)
temperature. At low temperature the sample exhibits a typical
''terminal'' behavior of a Maxwell fluid, $G'' = \eta_0 \omega$ with
an effective viscosity $\eta_0 =0.32$ Pa s, angular frequency
$\omega = 2 \pi f$, and $G'=  \eta_0 \tau \omega^2$, with a
relaxation time $\tau= 0.056$ s. At room temperature, the behavior
is typical of a viscoelastic solid, with a frequency-independent
storage modulus more than one order of magnitude larger than the
weakly frequency-dependent loss modulus. Note that the transition is
reversible, as the subsequent cooling of the viscoelastic solid
leads back to the initial low-temperature Maxwell fluid.

\begin{figure}
\includegraphics[width=0.45\textwidth]{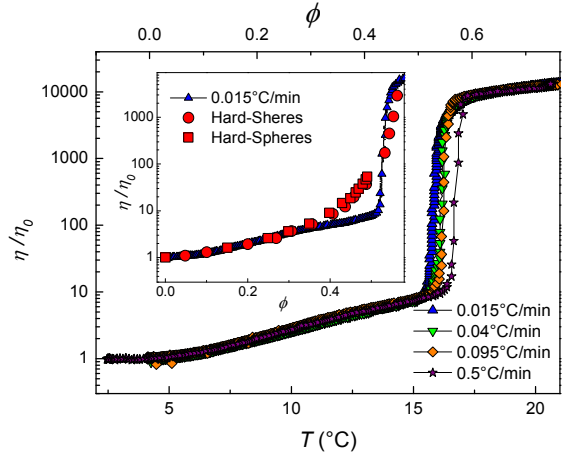}
\caption{(Color online) Complex viscosity normalized by the viscosity of the
solvent at low temperature ($0.32$ Pa.s) as a function of
temperature (bottom axis) and effective volume fraction (top axis),
for samples prepared with different heating rates as indicated in
the legend. Inset: Viscosity of the suspension normalized by the viscosity of the solvent versus volume fraction for
our samples (blue triangles) and for two independent measurements
\cite{Segre95,Cheng02} of colloidal hard spheres suspensions (red
symbols). } \label{fig:ViscosityvsT}
\end{figure}

In order to rationalize our experimental results, we propose to use
existing theories for colloidal suspensions, for which the standard
control parameter is the volume fraction of colloids. To this aim,
the first step is to provide a mapping between temperature and
volume fraction for the sample investigated here. A mapping has been derived in~\cite{MortensenPRL92}
by modeling the structure factor of
the micelles with that of hard spheres. For a system very similar to ours, the authors found
that the volume fraction of micelles varies linearly with temperature. Therefore, we assume
that the volume fraction of the micelles varies
linearly with $T$,
%+++++++++++++++++++++++++++++++++++++++++++
\begin{equation}
\phi = \alpha (T-T_0) \, ,\label{eq:Phi}
\end{equation}
%+++++++++++++++++++++++++++++++++++++++++++
and we propose \cite{Trong08} to determine the parameters $T_0$ and $\alpha$ by comparing our
rheology data to viscosity
measurements for colloidal hard-sphere suspensions~\cite{Segre95,Cheng02}.
The complex viscosity, $\eta=\frac{1}{2\pi f}\sqrt{G'^2 + G''^2}$,
is calculated and normalized by the complex viscosity at low
temperature ($\eta_0=0.32$ Pa.s) and plotted as a function of
temperature in fig.~\ref{fig:ViscosityvsT}. Up to about $\sim
4.2~{^\circ}\mathrm{C}$, the viscosity is constant, and then it
increases smoothly up to a temperature around
$15~{^\circ}\mathrm{C}$, where it has grown by one order of
magnitude. We find the best collapse of
the two sets of data for $T_0=4.2~{^\circ}\mathrm{C}$ and
$\alpha=0.0455~{^\circ}\mathrm{C}^{-1}$
(fig.~\ref{fig:ViscosityvsT}). A reasonable agreement is found up to
a $5$ fold increase of the viscosity corresponding to a volume
fraction of about $0.35$. For volume fractions in the range
$0.35-0.50$ (before crystallization), the reduced viscosity of the
copolymer sample increases more gently than that of supercooled hard
spheres suspensions. This discrepancy may have several origins:
first of all, as micelles form, the ``solvent viscosity'' might
decrease due to polymer consumption. Moreover, the aggregation
number and the size of the micelles might slightly increase with
temperature~\cite{MortensenEPL92}, inducing slight changes in the effective
volume fraction. Finally, the micelle softness, as compared to hard
sphere particles, might play a role. Indeed, it is known that at low  $\phi$
the viscosity of soft and hard spheres behave very similarly, while at
larger $\phi$  soft spheres are less viscous than a corresponding suspension
of hard particles (see e.g. \cite{SessomsPTRSA2009,Merlet10}). In spite of these discrepancies
at $\phi > 0.35$, we find that crystallization occurs for an
equivalent hard sphere volume fraction on the order of $0.52-0.56$,
a range similar to that observed experimentally for hard-sphere
suspensions.

The temperature dependence of the complex modulus
(fig.~\ref{fig:1}a) can be rationalized using the mapping thus
established:  from $T=T_0$ to $T=T_c$, the sample becomes more and
more viscous ($G''$ increases) as the volume fraction of micelles
increases continuously. The sample elasticity is not affected ($G'$
remains small and constant), until crystallization occurs due to
micelle crowding, yielding the abrupt upward jump of both $G'$ and
$G''$.

%-----------------------------------  FIG2 -------------------------------------
%------------------------------------ FIG2 --------------------------------------

\subsection{Effect of the heating rate on the crystallization temperature}

\begin{figure}
\includegraphics[width=0.45\textwidth]{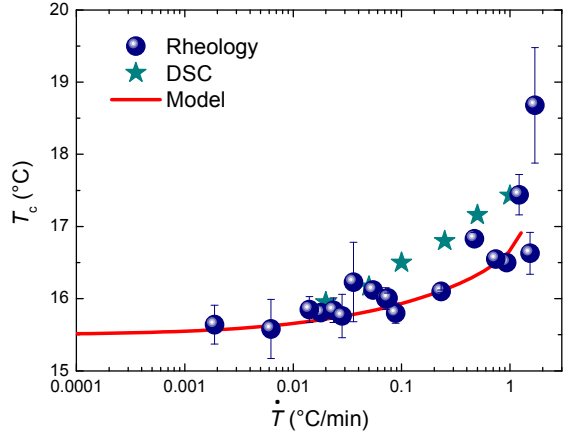}
\caption{(Color online) Crystallization temperature as a function of the heating
rate as measured by linear rheology (blue circles) and DSC (green
stars). The red line is a fit to the experimental data using the
model described in the text.} \label{fig:TcvsTdot}
\end{figure}

Visco-elasticity data measured for different heating rates are shown
in figs.~\ref{fig:1}b and~\ref{fig:ViscosityvsT}. Prior to
crystallization, all data collapse nicely, indicating that the rate
of increase of the micelle volume fraction is not controlled by
kinetics factors but is instead controlled by thermodynamics and
thus depends only on temperature. Similarly, data collapse in the
solid crystalline phase at high temperature, and hence high volume fraction, showing
that the mechanical properties of the solid phase are robust with
respect to a change of the solidification protocol. Note however
that there are slight differences in the loss modulus, which will be
analyzed in details elsewhere. By contrast, we find that the
crystallization temperature depends significantly on $\dot{T}$.
Although this is not a priori surprising since crystallization
involves kinetic processes, it is interesting to observe that simple
rheology experiments can capture these effects. Figures~\ref{fig:1}b
and~\ref{fig:ViscosityvsT} show that the crystallization temperature
is shifted toward higher temperatures as $\dot{T}$ increases. The
evolution with the heating rate of the crystallization temperature,
$T_c$, as measured by rheology, is plotted in
fig.~\ref{fig:TcvsTdot}, for $\dot{T}$ spanning  more than three
orders of magnitude. Numerical values are in the range
$(15.5-19.5)~{^\circ}\mathrm{C}$. Although rather scattered, data
clearly indicate a decrease of $T_c$ when $\dot{T}$ decreases
leading eventually to an ``equilibrium'' value $T_c \simeq
15.5~{^\circ}\mathrm{C}$ for very slow heating rates, essentially
independent of $\dot{T}$.

\begin{figure}
\includegraphics[width=0.45\textwidth]{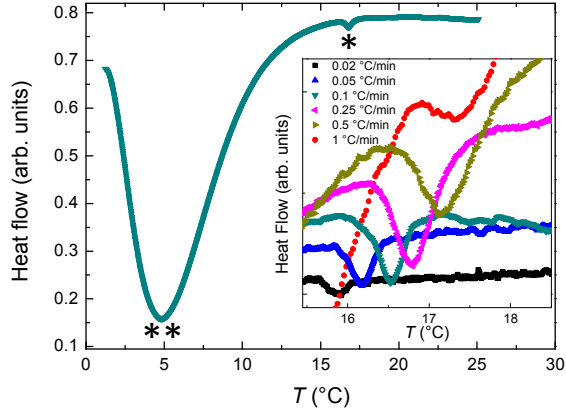}
\caption{(Color online) Heat flow versus temperature when temperature is varied
from $0.5$ to $25~{^\circ}\mathrm{C}$ at a rate of
$0.1~{^\circ}\mathrm{C/min}$. Inset: zoom of the data in the zone
when the crystallization peak occurs, for data acquired at different
temperature rates as indicated in the legend. Data in the inset have
been arbitrarily shifted along the y axis. ** and * point
respectively to the peaks due to micellization and crystallization.}
\label{fig:DSC}
\end{figure}

We complement the rheology data with differential
scanning calorimetry (DSC) measurements of the
crystallization temperature. A complete DSC curve is shown in
fig.~\ref{fig:DSC}. One can distinguish a very broad and deep
inverse peak, due to micellization, which is centered around
$T=5~{^\circ}\mathrm{C}$. This peak is followed by a low amplitude
peak around $16~{^\circ}\mathrm{C}$, which is due to the
crystallization of the micelles~\cite{Trong08,Lau04}. DSC data acquired at different $\dot{T}$, although
in a range smaller than for rheology experiments due to experimental
limitations, show a similar shift toward higher temperature of
crystallization as $\dot{T}$ increases. Numerical values of the
crystallization temperature extracted from DSC are plotted together
with the rheology data in fig.~\ref{fig:TcvsTdot}. Data measured
with the two techniques are compatible, supporting the method used
to extract a crystallization temperature from visco-elasticity
measurements.

%-----------------------------------  FIG3 -------------------------------------
%------------------------------------ FIG3 --------------------------------------

\section{Modeling the Nucleation and Growth of Crystals}
\label{Sec:Model}

The fact that crystallization occurs at a higher temperature (higher
$\phi$) when solidification proceeds faster can be understood by
recalling that crystallization is a kinetic process. As shown in a
so-called time-temperature-transformation diagram~\cite{bookKurz}
(TTT diagram), crystallization occurs at small supercooling (smaller
volume fraction, and hence lower $T$ in our case) for a very slow
temperature ramp, while it occurs at higher supercooling (larger
volume fraction, higher $T$) for very fast ramps. To quantitatively
account for our experimental observations, however, one needs to
take into account not only the kinetically controlled nucleation
process but also the growth process of the crystalline phase. We use
here standard theories for the nucleation and growth of crystalline
phases, initially developed for atomic and molecular systems and
subsequently applied to colloidal suspensions. Moreover, we adapt
these theories in order to take into account the presence of
nanoparticles and the non-isothermal conditions under which the
sample solidifies.

%-----------------------------------  FIG4 -------------------------------------
%------------------------------------ FIG4 --------------------------------------

\subsection{Crystallization temperature and average crystallite size}

We first recall how an average grain size can be computed, using the
Johnson-Mehl-Avrami-Kolmogorov (JMAK) theory
\cite{Kolmogorov37,Johnson39,Avrami39,Avrami40,Avrami41}. In this
theory one assumes that nucleation occurs randomly and homogeneously
and that the growth speed of the crystallites, $v_G$, does not depend
on the extent of the crystallization process and is isotropic.
Following JMAK, the extended crystal fraction at time $t$,
$X_e(t)$, that is the volume fraction of the whole sample which is occupied by crystals, reads:

%+++++++++++++++++++++++++++++++++++++++++++
\begin{equation}
X_{e}(t) =  \frac{4 \pi}{3} \int_{0}^{t}  I(\tau)  \left [
\int_{\tau}^{t}v_G(\tau')\mathrm{d}\tau'\right]^3  \mathrm{d}\tau
\,,\label{eq:X_ext}
\end{equation}
%+++++++++++++++++++++++++++++++++++++++++++
where $I(\tau)$ is the nucleation rate per unit volume at time
$\tau$, and  $ \frac{4 \pi}{3} \left [
\int_{\tau}^{t}v_G(\tau')\mathrm{d}\tau'\right]^3$ is the volume at
time $t$ of a grain that has nucleated at time $\tau$ and has grown
from time $\tau$ to time $t$, with a time-dependent growth velocity
$v_G(\tau')$.

The extended crystal fraction does not take into account the
impingement between crystallites. As shown by Kolmogorov and Avrami,
the actual crystal fraction is
related to $X_e(t)$ by

%+++++++++++++++++++++++++++++++++++++++++++
\begin{equation}
X(t) =  1-\exp[-X_e(t)]
\label{eq:JMAK}
\end{equation}
%+++++++++++++++++++++++++++++++++++++++++++

The average grain size $R$ at the end of the crystallization process
can be derived from the final grain
density~\cite{Farjas07,Farjas08}:
%+++++++++++++++++++++++++++++++++++++++++++
\begin{equation}
R =  \left ( \frac{3}{4\pi} \int_{0}^{\infty}  I_a(\tau) \mathrm{d}\tau \right ) ^{-1/3}
\label{eq:R}
\end{equation}
%+++++++++++++++++++++++++++++++++++++++++++
Here $I_a$ is the actual nucleation rate which takes into account
the fact that nucleation can only occur in regions which have not
crystallized yet:
%+++++++++++++++++++++++++++++++++++++++++++
\begin{equation}
I_a(\tau) =  \left [ 1- X(\tau)\right ] I(\tau) \,.\label{eq:I_a}
\end{equation}
%+++++++++++++++++++++++++++++++++++++++++++

%-----------------------------------  FIG5 -------------------------------------
\begin{figure}
\includegraphics[width=0.40\textwidth]{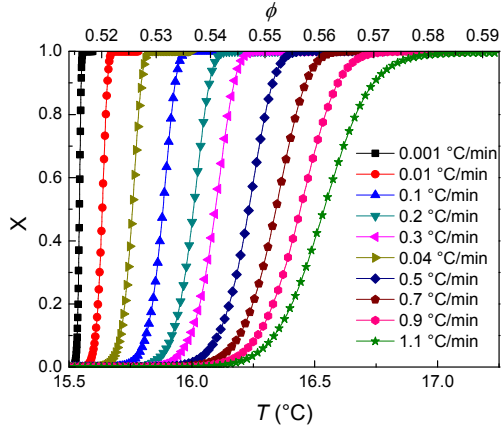}
\caption{(Color online) Evolution of the crystal fraction $X$, with temperature (bottom $x$-axis) and with volume fraction (top $x$-axis). $X$ is computed with the same parameters as the ones used to fit the experimental data (fig.~\ref{fig:TcvsTdot}). Data are shown for several heating rates, as indicated in the caption. }
\label{fig:VolumFract}
\end{figure}
%------------------------------------ FIG5 --------------------------------------

The above equations are very general and hold for time-dependent
nucleation and growth rates, as is the case in our experiments where
a continuous heating is imposed. Note that, using these equations, the standard result obtained in the case of
time-independent nucleation and growth rates can be recovered~\cite{Auer05}. In this case, $ X_{e}(t) = \frac{4
\pi}{3} I v_G^3 t^4 $. If one moreover neglects impingement and
considers the extended volume fraction of crystallized material as
the actual volume fraction, the time $t^*$ at which crystallization
stops reads $t^* \sim (I v_G^3)^{-1/4}$. Hence the average
crystallite size (eq.~\ref{eq:R}) reads $R \sim  I \left (
\int_{0}^{t^*}  \mathrm{d}\tau \right ) ^{-1/3}  \sim \left (
\frac{v_G}{I} \right )^{1/4}$, as previously derived but not
experimentally verified for colloidal samples~\cite{Auer05}.

For a time-varying process, one can use Eqs.~(\ref{eq:X_ext}) and (\ref{eq:JMAK}) to calculate the temporal evolution of the crystallinity, provided that expressions for $v_G$ and $I$ are given. Before discussing in detail how to model the $T$- (or, equivalently, $\phi$-) dependence of the nucleation and growth rates, we anticipate the results for the calculated $X(t)$ in the case of our samples. The evolution of the actual volume fraction of the sample that has crystallized is shown as a function of temperature (bottom axis) or volume fraction (top axis) in
fig.~\ref{fig:VolumFract}, for several $\dot{T}$ spanning three
orders of magnitude. We take as time $t=0$ the time at which the
sample temperature reaches the equilibrium solidification temperature
($T=15.5~{^\circ}\mathrm{C}$, $\phi=0.514$). All curves have
sigmoidal shapes and reach asymptotically $1$. They are shifted towards higher temperatures(and hence higher volume fractions) when $\dot{T}$
increases, as observed experimentally. Although $X(t)$ reaches $1$
only asymptotically, the curves shown in fig.~\ref{fig:VolumFract} can be
used to define a crystallization time, $t^*$, by setting a threshold $X_{\rm th}$ such that $X(t^*)= X_{\rm th}$. Knowing the temperature history
imposed to the sample, the evaluation of temperature $T^*$ at which
crystallization is completed is then straightforward. Having established how the crystallization temperature can be in principle calculated, we now need to provide analytical expressions for the nucleation and growth rates.

\subsection{Nucleation rate, growth rate, and the effect of nanoparticles}
\label{subsec:vGI}

We assume that crystallites grow into spherical particles at a rate $v_G$, which is given by the Wilson-Frenkel equation~\cite{Auer05}:
%+++++++++++++++++++++++++++++++++++++++++++
\begin{equation}
v_G=\frac{D_s}{\lambda}[1-\exp(-|\Delta\mu|/k_BT)] \,.
\label{eq:Vg}
\end{equation}
%+++++++++++++++++++++++++++++++++++++++++++++++
Here $D_s$ is the long-time self diffusion coefficient of the particles (atoms, molecules or colloids) in the fluid phase, $\lambda$ is the typical distance over which a particle diffuses to become part of a growing crystallite, $\Delta\mu$ is the difference in chemical potential between the solid and the fluid phases, and $k_B$ is Boltzmann's constant. Note that $\lambda$ is expected to be of the order of the particle size. All quantities depend on temperature for atomic and molecular systems, and on volume fraction for colloidal systems.

We use the classical nucleation theory (CNT) to evaluate the nucleation rate. Classical nucleation theory predicts that the nucleation rate per unit volume, $I$, is an Arrhenius-like function \cite{bookKelton}:
%+++++++++++++++++++++++++++++++++++++++++++
\begin{equation}
I=\Gamma \exp [-\Delta G^*/k_BT] \,,
\label{eq:I}
\end{equation}
%+++++++++++++++++++++++++++++++++++++++++++++++
where $\Delta G^*$ is the nucleation barrier and $\Gamma$ the kinetic factor~\cite{Auer05}:
%+++++++++++++++++++++++++++++++++++++++++++
\begin{equation}
\Gamma = \left (\frac{1}{6 \pi k_BT n_c}|\Delta\mu|\right)^{0.5} \rho_L \frac{24 D_s}{\lambda ^2} n_c^{2/3} \,.
\label{eq:Gamma}
\end{equation}
%+++++++++++++++++++++++++++++++++++++++++++++++
In Eq.~(\ref{eq:Gamma}), $n_c$ is the number of particles in a critical nucleus and $\rho_L$ is the particle number density in the fluid phase.
In CNT, the critical radius and the nucleation barrier are calculated by balancing the surface Gibbs free energy of a crystalline nucleus of radius $r$, $E_S= 4 \pi r^2\gamma_0 $  and the volume Gibbs free energy term,  $E_V= - \frac{4 \pi}{3}  \rho_S |\Delta \mu| r^3$, where $\gamma_0$ is the surface energy between the fluid and the crystal phase and $\rho_S$ is the number density of the particles in the solid phase. Since $E_s$ grows as $r^2$ while $E_V$ decrease as $-r^3$, the net change of free energy for a nucleus as compared to the fluid phase goes through a maximum: the free energy barrier to be overcome for a nucleus to be stable is then $\Delta G^* = \frac{16 \pi}{3} \frac{\gamma_0^3}{(\rho_S |\Delta \mu|)^2}$.

We now provide a model to account for the presence of a small amount of NPs of diameter $\sigma_P$ that are added to the copolymer suspension, at a volume fraction $\phi_P$. We have previously observed that upon solidification the NPs concentrate in the grain-boundaries and influence the microstructure of the polycrystals \cite{Ghofraniha12}. In the following, we neglect the effect of the diffusion of NPs on crystal growth, which could occur because of NPs partitioning, and thus assume that $v_G$ remains unchanged (Eq.~(\ref{eq:Vg})). By contrast, we argue that the NPs influence the barrier for nucleation, $\Delta G^*$, because they are trapped at the interface between the crystalline phase and the fluid phase, as observed experimentally \cite{Ghofraniha12,Zhang05}. We assume that the net effect of this trapping is a decrease of the surface free energy of the crystalline nuclei. To model this phenomenon, we adapt the CNT calculation of the free energy barrier by modifying the surface free energy term $E_S$. We propose that the surface free energy is reduced by an amount $E'_S$ proportional to the surface occupied by the NPs trapped at the
interface. Since these particles were initially in the volume occupied by the crystallite (the number of particles in a crystallite of radius $r$ is $8 r^3\phi_P /\sigma_P^3$), their number scales as $r^3\phi_P /\sigma_P^3$. Thus, $E'_S \sim \sigma_P^2 r^3\phi_P /\sigma_P^3$, assuming that each NP at the interface contributes to a reduction of surface energy proportional to the area it occupied at the interface, $\frac{\pi}{4} \sigma_P^2$, so that the modified surface free energy reads:
%+++++++++++++++++++++++++++++++++++++++++++
\begin{equation}
E_S= 4 \pi\gamma_0   r^2 - E'_S= 4 \pi \gamma_0 r^2  - \gamma_0 \frac{2 \pi E_{P} \phi_P} {\sigma_P} r^3 \,
\label{eq:ES}
\end{equation}
%+++++++++++++++++++++++++++++++++++++++++++++++
where $E_P$ is a proportionality constant accounting for both the partitioning of the NPs (i.e. for what fraction of them is actually rejected at the interface of the growing crystallite) and the reduction of surface tension due to one single NP. Note that the fact that impurities could affect the liquid-solid surface tension was already mentioned in Ref.~\cite{Lee99} in the context of metallic alloys and in Ref.~\cite{Miller87} in the context of phospholipid monolayers.

By balancing the surface and volume terms, we find a modified equation for the nucleation barrier:
%+++++++++++++++++++++++++++++++++++++++++++
\begin{equation}
\Delta G^* = \frac{16 \pi}{3} \frac{\gamma_0^3}{[\rho_S |\Delta \mu|  + \frac{3}{2} \gamma_0 E_{P} \phi_P / \sigma_P]^2} \,.
\label{eq:DeltaG*}
\end{equation}
%+++++++++++++++++++++++++++++++++++++++++++++++
Interestingly, the NP contribution, although physically
due to a change of the surface term, can formally be incorporated
in the volume term $E_V$, because it scales with the volume of the nucleus. The net effect of the NPs is thus to lower the nucleation barrier thanks to an effective increase of the volume term.

\section{Analysis of the Experimental Data}
\label{Sec:Analysis}

\subsection{Numerical values of the parameters}

In this subsection, we detail the numerical values and the fitting parameters used in evaluating $v_G$ and $I$ as defined in Sec.~\ref{subsec:vGI}. Several parameters can be evaluated from previous numerical and experimental works on hard-sphere colloidal suspensions, finally leaving four fitting parameters. In all calculations, we use the mapping between volume fraction and temperature determined experimentally, Eq.~(\ref{eq:Phi}), assuming that this mapping is valid in the whole range of volume fractions considered. The number density in the liquid phase is $\rho_L =  \frac{6  \phi}{\pi \sigma^3}$, with $\sigma=22$ nm the micelle diameter. To reduce the number of fitting parameters with no significant loss of generality, we assume a constant volume fraction in the solid phase ($\phi_s=0.5$), so that $\rho_S =  \frac{6  \phi_s}{\pi \sigma^3}$. We furthermore fix $n_c=100$, since for hard spheres the number of particles in a critical nucleus has been found numerically to depend weakly on volume fraction~\cite{Auer05}. Note that the exact value of $n_c$ does not have a strong influence on the calculations, since only the kinetic parameter $\Gamma$ depends on $n_c$, but very weakly, $\Gamma \sim n_c^{1/6}$ (Eq.~\ref{eq:I}). Finally, the micelle diffusion coefficient $D_s$ is strongly dependent on volume fraction, and hence on temperature. $D_s$ could be in principle estimated from the viscosity data of Fig.~\ref{fig:ViscosityvsT}. However, viscosity is measured macroscopically on samples where crystalline and fluid phases may coexist, while Eqs.~(\ref{eq:Vg}) and (\ref{eq:Gamma}) require the microscopic diffusion coefficient in the fluid phase alone. We thus estimate $D_s$ from measurements of the microscopic structural relaxation time, $\tau_{\alpha}$, of supercooled suspensions of colloidal hard spheres, assuming $\frac{D_s}{D_{s,0}} = \frac{\tau_{\alpha,0}}{\tau_{\alpha}}$, where the index $0$ refers to quantities in the limit $\phi \rightarrow 0$. By fitting the data of Ref.~\cite{BrambillaPRL2009} for $\phi \leq 0.575$, we find the empirical law $D_s=D_0 \left [ (1-2.5\phi+1.36\phi^2)(1-\frac{\phi}{\phi_c})+11.056 \phi^2 (1-\frac{\phi}{\phi_c})^{2.5}  \right ]$, with $\phi_c=0.598$ and $D_0 = 6 \times 10^{-14} \rm{m}^2 \rm{s}^{-1}$ evaluated from the Stokes-Einstein relation, $ D_s = \frac{k_B T}{3 \pi \eta_0 \sigma}$, using $\eta_0=0.32\rm{Pa~s}$. This empirical law holds to a very good approximation up to $\phi \sim 0.55$ and is consistent with our viscosity data for the micellar suspension up to the onset of crystallization. Although at higher $\phi$  this law eventually departs from the data of Ref.~\cite{BrambillaPRL2009}, we have checked that the detailed form of $ D_s$  at very high $\phi$  is irrelevant for our findings. 

The remaining parameters appearing in $v_G$ and $I$ are treated as fitting parameters. According to numerical work on hard spheres~\cite{Auer05}, the volume fraction dependence of the chemical potential difference between the solid and the liquid can be well adjusted by the simple functional form
$\Delta \mu = A_{\mu} k_BT (\phi_{crys} - \phi) $, with $\phi_{crys} = 0.49776$ and $A_{\mu}=14.7$. Here we consider the prefactor $A_{\mu}$ as a fitting parameter and take $\phi_{crys} = 0.514$. With this choice, $\Delta \mu = 0$ for $T=15.5 ~{^\circ}\mathrm{C}$, which is the experimental numerical value of $T_c$ measured for extremely slow heating rates (see fig.~\ref{fig:TcvsTdot}), for which the crystallization temperature is nearly independent of the heating rate. The surface tension between the fluid and solid phases in colloidal suspension has been measured experimentally by several groups. They all find $\gamma_0 = k_BT  \frac{A_{\gamma}}{\sigma^2}$ with $A_{\gamma}$ of order one~\cite{HernandezGuzman09,Ramsteiner10,Nguyen11}. In the following, we use the same expression taking $A_{\gamma}$ as a fitting parameter. Numerical simulations and experiments show that the typical distance over which diffusion occurs is proportional to the colloid size:  $\lambda = A_{\lambda}\sigma $, although there is  a significant discrepancy between simulation and experiments: $A_{\lambda}$ is found in the range $0.3-0.5$ in simulations~\cite{Auer05}, while it ranges from $3$ to $17$ in experiments~\cite{Palberg99}.  In the following, we assume $\lambda = A_{\lambda}\sigma $ with $A_{\lambda}$ a fitting parameter. The fourth adjusting parameter is $E_p$, the proportionality constant that accounts for the effect of NPs in the reduction of the surface tension already discussed in Sec.~\ref{subsec:vGI}, see Eq.~(\ref{eq:DeltaG*}).

\subsection{Fit of the experimental data}

We use the model developed above to fit the experimental crystallization temperature as a function of temperature increase rate (fig.~\ref{fig:TcvsTdot}). We remind that rheology measurements yield $T_c$, defined as the temperature at which the sample response becomes predominantly solid-like (the storage modulus $G'$ becomes larger than the loss modulus $G''$). In the model, one needs to define a threshold for the crystallinity $X$ in order to define the crystallization temperature, $T^*$, since the sample is fully crystallized only for $t\rightarrow \infty$ (fig.~\ref{fig:VolumFract}). We identify $T_c$ with $T^*$ choosing a threshold $X_{\rm th} = 0.9$ and check that the results are essentially threshold-independent for $0.5 \le X_{\rm th} \le 0.98$. The best fit to the experiments is shown as a continuous line in fig.~\ref{fig:TcvsTdot}, showing a good agreement with the data. This adjustment was obtained with the following values of the fitting parameters: $A_{\mu} = 10 \pm 2$, $A_{\gamma} = 1.5 \pm 0.1$, $A_{\lambda} = 0.11 \pm 0.3$, $E_P = 113 \pm 5$. The values for $A_{\mu}$, $A_{\gamma}$ and $A_{\lambda}$ are in agreement with those expected from numerical simulations and experiments on hard-sphere colloidal suspensions  (see Sec.~\ref{Sec:Analysis}). They are also reasonably close to the values extracted using the same model for reproducing the crystallite size as a function of NP content and temperature rate~\cite{Ghofraniha12} [$A_{\mu} = 17 \pm 3$, $A_{\gamma} = 0.75 \pm 0.2$, $A_{\lambda} = 0.14 \pm 0.05$, $E_P = 126 \pm 24$], albeit with a different kind of nanoparticle impurities.

%-----------------------------------  FIG6 -------------------------------------
\begin{figure}
\includegraphics[width=0.40\textwidth]{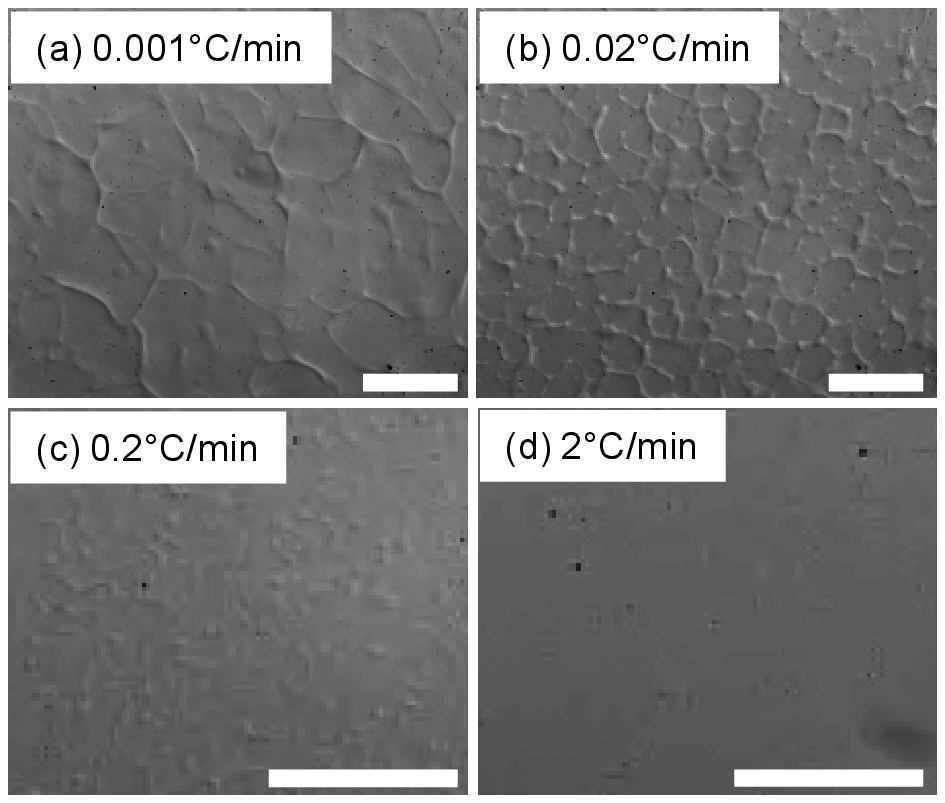}
\caption{Differential Interference Contrast light microscopy of the samples prepared with different heating rates as indicated in the legend. Scale bars: $20 \mu$m}
\label{fig:microscopy}
\end{figure}
%------------------------------------ FIG6 --------------------------------------

\begin{figure}
\includegraphics[width=0.40\textwidth]{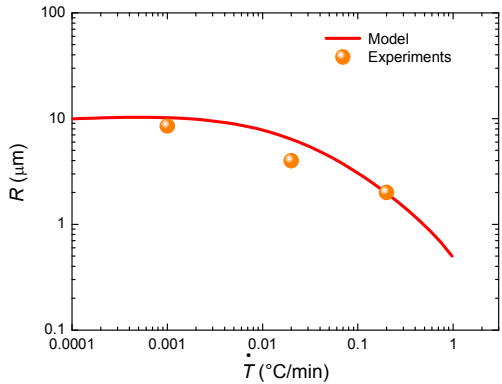}
\caption{(Color online) Average size of the crystallites as a function of the heating rate. The symbols correspond to the experimental data and the line is the numerical values calculated with the model detailed in the text. The adjusting parameters are the ones extracted from the fit of the experimental data of fig.~\ref{fig:TcvsTdot}.}
\label{fig:Radius}
\end{figure}

As a further check of the soundness of our approach, we compare the $\dot{T}$ dependence of the average crystallite size obtained from the model to that measured experimentally by analyzing images of the polycrystalline microstructure (fig.~\ref{fig:microscopy}). The images are taken by differential interference contrast optical microscopy and the microstructure is visible thanks to the presence of nanoparticles that segregate in the grain-boundaries, thereby providing the required optical contrast between adjacent grains. To calculate the average grain size, $R$, as a function of $\dot{T}$, we use Eq.~(\ref{eq:R}) with the same set of fitting parameters as determined by fitting $T_c(\dot{T})$. Experimental and numerical results are shown in fig.~\ref{fig:Radius} as symbols and a line, respectively. For $\dot{T}$ in the range $10^{-4}-1 \Cmin$, the model predicts $R$ to be in the micrometer range and to continuously decrease as $\dot{T}$ increases, as observed previously~\cite{Ghofraniha12}. Note that the drop of $R$ is significant at relatively fast heating rates, while the grain size levels off at about $10\um$  for very slow rates. Indeed, as discussed in Ref.~\cite{Ghofraniha12}, for slow ramps the crystallite size is only limited by the presence of nanoparticles. The model is in very good agreement with the experimental values of $R$ for samples prepared with temperature rates in the range $0.001-0.2\Cmin$, capturing both the decreasing trend of $R$ and its absolute value, thus providing additional support to our analysis. Interestingly, the grain size can be measured easily for the lowest heating rates, for which the network of grain boundaries is clearly visible in the microscope images (fig.~\ref{fig:microscopy}a, b). For $\dot{T} = 0.2 \Cmin$ (fig.~\ref{fig:microscopy}c) the contrast is poorer, but the grain boundaries are still visible. By contrast, the sample prepared with the fastest heating rate, $\dot{T} = 2 \Cmin$ (fig.~\ref{fig:microscopy}d), appears uniform, presumably due to the very small grain size, as predicted by the model, and to a less efficient partitioning of the NP between the bulk and the grain boundaries, as also observed in Ref.~\cite{Ghofraniha12}.

%-----------------------------------  FIG7 -------------------------------------
%------------------------------------ FIG7 --------------------------------------

\subsection{Discussion}

We have used a thermosensitive micellar system, for which the volume micelle volume fraction can be tuned with temperature, to investigate the crystallization dynamics of colloidal suspensions under time-varying volume fraction conditions. The crystallization temperature $T_c$ was measured using standard rheometry and its dependence on the rate of the temperature ramp used to solidify the sample was investigated. The experimental data can be very well accounted for by using a standard model for the nucleation and growth of colloidal crystals that has been adapted to the case of time-varying volume fraction conditions and to the presence of NP impurities. Our model predicts also how the microstructure of the polycrystal evolves with $\dot{T}$, which is found in good quantitative agreement with the experiments, thus providing a successful cross-check of our approach. We have therefore demonstrated that rheometry experiments combined with standard models for the nucleation and growth of colloidal crystals allow quantitative parameters related to the crystallization processes to be derived.

\begin{figure}
\includegraphics[width=0.45\textwidth]{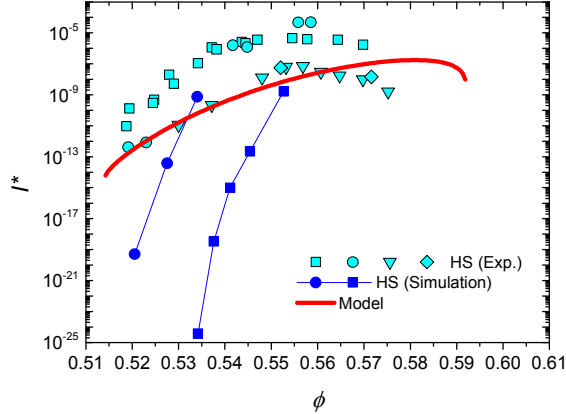}
\caption{(Color online) Normalized nucleation rate, $I^* = I\sigma^5 /D_0$, as a function of particle volume fraction. Symbols are experimental data for hard spheres suspensions of Ref.~\cite{Sinn01} (squares), Ref.~\cite{Schatzel93} (circles), Ref.~\cite{Cheng02} (diamonds), Ref.~\cite{Harland97} (triangles). Line and symbols are numerical simulations of monodisperse (circles) and $5 \%$ polydisperse (squares) hard sphere suspensions, taken from Ref.~\cite{Auer05}. The thick line is our model. Figure adapted from~\cite{Auer05}.}
\label{fig:Irate}
\end{figure}

Our analysis was performed assuming that the behavior of the suspension of copolymer micelles can be mapped to that of hard-sphere suspensions. To proceed a step further along this analogy, we compute the nucleation rate, as given by Eqs.~(\ref{eq:I}) and~(\ref{eq:DeltaG*}), using the set of adjusting parameters determined in Sec.~\ref{Sec:Analysis}, and compare it to results from experiments and simulations of hard spheres. Data taken from the literature are collected and plotted as a function of volume fraction in fig.~\ref{fig:Irate}, together with our model. In order to compare data obtained with different particles and solvent viscosity, we plot the normalized nucleation rate $I^*=I \sigma^5 /D_0$  where $\sigma$ is the colloid diameter and $D_0$ its diffusion coefficient in the dilute regime~\cite{Auer05}. Interestingly, one finds a good agreement between the experimental data of hard sphere suspensions and the curve extracted from our model, the agreement being much less good with results from numerical simulations of hard sphere suspensions. More quantitatively, our data are slightly shifted towards higher volume fraction as compared to data for hard-spheres. We indeed find that the maximum of $I^*$  occurs at $\phi=0.58$ for copolymer micelles, to be compared to $\phi=0.56$ for hard-sphere colloids. By contrast, the height of the maximum is of the same order of magnitude: we calculate a maximum normalized nucleation rate of $\sim 2 \, 10^{-8}$ for our micelles, whereas experiments on hard-spheres yield values in the range $2 \, 10^{-5}-7 \, 10^{-8}$. Despite small discrepancies, the remarkable agreement found here hints at profound analogies between hard-sphere suspensions and suspensions of Pluronics micelles, in spite of the different nature of the materials. In particular, the interaction potential between micelles is expected to be much softer than that between hard-spheres. The softness of the potential is known to have a profound influence on the suspension properties near jamming. However, this appears not to be the case when the packing fraction is still well below random close packing, as at the onset of crystallization, presumably because the particles hardly experience any direct contact.

To conclude, thanks to our approach combining experiments and modeling, we have
pointed out analogies between hard-sphere colloidal crystals and Pluronics micellar crystals,
in spite of the difference in particle softness. This observation should be of particular interest to the numerous groups currently working on soft, thermoswallable microgel particles. In addition, we have provided a general framework
to rationalize the effect of temperature history or, more generally, volume fraction history, on colloidal crystallization. These results should be relevant for colloidal systems where volume fraction can be tuned, and in particular for the vastly investigated Pluronics
gels. Finally, we have demonstrated that crystallization processes can be
quantitatively probed using standard rheometry, an experimental method more readily accessible than most microscopic characterization tools, such as neutron or X-ray scattering.

%-----------------------------------  FIG8 -------------------------------------
%------------------------------------ FIG8 --------------------------------------

\section{Acknowldegments}
We thank  Peter Olmsted for illuminating discussions on the modeling of the influence of NPs on the nucleation rate. This work has been supported by ANR
under Contract No. ANR-09-BLAN-0198 (COMET).

%--------------------------------------------- REFERENCES -----------------------------------------

%\bibliography{biblio-grain}

\begin{thebibliography}{33}
\expandafter\ifx\csname natexlab\endcsname\relax\def\natexlab#1{#1}\fi
\expandafter\ifx\csname bibnamefont\endcsname\relax
  \def\bibnamefont#1{#1}\fi
\expandafter\ifx\csname bibfnamefont\endcsname\relax
  \def\bibfnamefont#1{#1}\fi
\expandafter\ifx\csname citenamefont\endcsname\relax
  \def\citenamefont#1{#1}\fi
\expandafter\ifx\csname url\endcsname\relax
  \def\url#1{\texttt{#1}}\fi
\expandafter\ifx\csname urlprefix\endcsname\relax\def\urlprefix{URL }\fi
\providecommand{\bibinfo}[2]{#2}
\providecommand{\eprint}[2][]{\url{#2}}




\bibitem{piazza00} R. Piazza, Current Opinion in Colloid and Interface Science,  \textbf{5}, 38, 2000.


\bibitem{pusey86} P. N. Pusey and W. Vanmegen, Nature,  \textbf{320}, 340, 1986.



\bibitem[{\citenamefont{Schatzel}(1993)\citenamefont{Schatzel}}]{Schatzel93}
\bibinfo{author}{\bibfnamefont{K.}~\bibnamefont{Sch\"atzel}},
   \bibnamefont{and} \bibinfo{author}{\bibfnamefont{B. J.} \bibnamefont{Ackerson}},
  \bibinfo{journal}{Phys. Rev. E} \textbf{\bibinfo{volume}{48}},
  \bibinfo{pages}{3766} (\bibinfo{year}{1993}).

\bibitem[{\citenamefont{Harland}(1997)\citenamefont{Harland}}]{Harland97}
\bibinfo{author}{\bibfnamefont{J. L.}~\bibnamefont{Harland}},
   \bibnamefont{and} \bibinfo{author}{\bibfnamefont{W.} \bibnamefont{van Megen}},
  \bibinfo{journal}{Phys. Rev. E} \textbf{\bibinfo{volume}{55}},
  \bibinfo{pages}{3054} (\bibinfo{year}{1997}).

\bibitem[{\citenamefont{Henderson}(1998)\citenamefont{Henderson}}]{Henderson98}
\bibinfo{author}{\bibfnamefont{S. I.}~\bibnamefont{Henderson}},
   \bibnamefont{and} \bibinfo{author}{\bibfnamefont{W.} \bibnamefont{van Megen}},
  \bibinfo{journal}{Phys. Rev. Lett.} \textbf{\bibinfo{volume}{80}},
  \bibinfo{pages}{877} (\bibinfo{year}{1998}).

  \bibitem[{\citenamefont{Palberg}(1999)\citenamefont{Palberg}}]{Palberg99}
\bibinfo{author}{\bibfnamefont{T.}~\bibnamefont{Palberg}},
  \bibinfo{journal}{J. Phys. C: Condens. Matter} \textbf{\bibinfo{volume}{11}},
  \bibinfo{pages}{R323} (\bibinfo{year}{1999}).

\bibitem[{\citenamefont{Sinn}(2001)\citenamefont{Sinn}}]{Sinn01}
\bibinfo{author}{\bibfnamefont{C.}~\bibnamefont{Sinn}},
\bibinfo{author}{\bibfnamefont{A.}~\bibnamefont{V}},
  \bibinfo{author}{\bibfnamefont{A.} \bibnamefont{Stipp}}, \bibnamefont{and}
  \bibinfo{author}{\bibfnamefont{T.} \bibnamefont{Palberg}},
  \bibinfo{journal}{Prog. Colloid Polym. Sci.} \textbf{\bibinfo{volume}{118}},
  \bibinfo{pages}{266} (\bibinfo{year}{2001}).

 \bibitem[{\citenamefont{Gasser}(2001)\citenamefont{Gasser}}]{Gasser01}
\bibinfo{author}{\bibfnamefont{U.}~\bibnamefont{Gasser}},
\bibinfo{author}{\bibfnamefont{E. R.}~\bibnamefont{Weeks}},
\bibinfo{author}{\bibfnamefont{A.}~\bibnamefont{Schofield}},
\bibinfo{author}{\bibfnamefont{P. N.}~\bibnamefont{Pusey}},
   \bibnamefont{and} \bibinfo{author}{\bibfnamefont{D. A.} \bibnamefont{Weitz}},
  \bibinfo{journal}{Science} \textbf{\bibinfo{volume}{292}},
  \bibinfo{pages}{258} (\bibinfo{year}{2001}).

\bibitem[{\citenamefont{Cheng}(2002)\citenamefont{Cheng}}]{Cheng01}
\bibinfo{author}{\bibfnamefont{Z.}~\bibnamefont{Cheng}},
\bibinfo{author}{\bibfnamefont{P. M.}~\bibnamefont{Chaikin}},
\bibinfo{author}{\bibfnamefont{J.}~\bibnamefont{Zhu}},
\bibinfo{author}{\bibfnamefont{W. B.}~\bibnamefont{Russel}},
   \bibnamefont{and} \bibinfo{author}{\bibfnamefont{M. V.} \bibnamefont{Meyer}},
  \bibinfo{journal}{Phys. Rev. Lett.} \textbf{\bibinfo{volume}{88}},
  \bibinfo{pages}{015501} (\bibinfo{year}{2001}).

 \bibitem[{\citenamefont{Martin}(2003)\citenamefont{Martin}}]{Martin03}
\bibinfo{author}{\bibfnamefont{S.}~\bibnamefont{Martin}},
\bibinfo{author}{\bibfnamefont{G.}~\bibnamefont{Bryant}},
   \bibnamefont{and} \bibinfo{author}{\bibfnamefont{W.} \bibnamefont{van Megen}},
  \bibinfo{journal}{Phys. Rev. E} \textbf{\bibinfo{volume}{67}},
  \bibinfo{pages}{061405} (\bibinfo{year}{2003}).

 \bibitem[{\citenamefont{Herlach}(2010)\citenamefont{Herlach}}]{Herlach10}
\bibinfo{author}{\bibfnamefont{D. M.}~\bibnamefont{Herlach}},
\bibinfo{author}{\bibfnamefont{I.}~\bibnamefont{Klassen}},
\bibinfo{author}{\bibfnamefont{P.}~\bibnamefont{Wette}},
   \bibnamefont{and} \bibinfo{author}{\bibfnamefont{D.} \bibnamefont{Holland-Moritz}},
  \bibinfo{journal}{J. Phys. C: Condens. Matter} \textbf{\bibinfo{volume}{22}},
  \bibinfo{pages}{153101} (\bibinfo{year}{2010}).

 \bibitem[{\citenamefont{Okubo}(2002)\citenamefont{Okubo}}]{Okubo02}
\bibinfo{author}{\bibfnamefont{T.}~\bibnamefont{Okubo}},
\bibinfo{author}{\bibfnamefont{H.}~\bibnamefont{Hase}},
\bibinfo{author}{\bibfnamefont{H.}~\bibnamefont{Kimura}},
   \bibnamefont{and} \bibinfo{author}{\bibfnamefont{E.} \bibnamefont{Kokufuta}},
  \bibinfo{journal}{Langmuir} \textbf{\bibinfo{volume}{18}},
  \bibinfo{pages}{6783} (\bibinfo{year}{2002}).

\bibitem[{\citenamefont{Meng}(2007)\citenamefont{Meng}}]{Meng07}
\bibinfo{author}{\bibfnamefont{Z.}~\bibnamefont{Meng}},
  \bibinfo{author}{\bibfnamefont{J. K.} \bibnamefont{Cho}},
    \bibinfo{author}{\bibfnamefont{S.} \bibnamefont{Debord}},
      \bibinfo{author}{\bibfnamefont{V.} \bibnamefont{Breedveld}},
  \bibnamefont{and} \bibinfo{author}{\bibfnamefont{L. A.} \bibnamefont{Lyon}},
  \bibinfo{journal}{J. Phys. Chem. B} \textbf{\bibinfo{volume}{111}},
  \bibinfo{pages}{6992} (\bibinfo{year}{2007}).

 \bibitem[{\citenamefont{Savage}(2009)\citenamefont{Savage}}]{Savage09}
\bibinfo{author}{\bibfnamefont{J. R.}~\bibnamefont{Savage}},
\bibnamefont{and} \bibinfo{author}{\bibfnamefont{A. D.} \bibnamefont{Dinsmore}},
  \bibinfo{journal}{Phys. Rev. Lett.} \textbf{\bibinfo{volume}{102}},
  \bibinfo{pages}{198302} (\bibinfo{year}{2009}).

 \bibitem[{\citenamefont{Palberg}(2009)\citenamefont{Palberg}}]{Palberg09}
\bibinfo{author}{\bibfnamefont{T.}~\bibnamefont{Palberg}},
\bibinfo{author}{\bibfnamefont{A.}~\bibnamefont{Stipp}},
   \bibnamefont{and} \bibinfo{author}{\bibfnamefont{E.} \bibnamefont{Bartsch}},
  \bibinfo{journal}{Phys. Rev. Lett.} \textbf{\bibinfo{volume}{102}},
  \bibinfo{pages}{038302} (\bibinfo{year}{2009}).

 \bibitem[{\citenamefont{Muluneh}(2012)\citenamefont{Muluneh}}]{Muluneh12}
\bibinfo{author}{\bibfnamefont{M.}~\bibnamefont{Muluneh}},
   \bibnamefont{and} \bibinfo{author}{\bibfnamefont{D. A.} \bibnamefont{Weitz}},
  \bibinfo{journal}{Phys. Rev. E} \textbf{\bibinfo{volume}{85}},
  \bibinfo{pages}{021405} (\bibinfo{year}{2012}).





\bibitem[{\citenamefont{Schope}(2006)\citenamefont{Schope}}]{Schope06}
\bibinfo{author}{\bibfnamefont{H. J.}~\bibnamefont{Sch\"ope}},
  \bibinfo{author}{\bibfnamefont{G.} \bibnamefont{Bryant}},
  \bibnamefont{and} \bibinfo{author}{\bibfnamefont{W.} \bibnamefont{van Megen}},
  \bibinfo{journal}{Phys. Rev. Lett.} \textbf{\bibinfo{volume}{96}},
  \bibinfo{pages}{175701} (\bibinfo{year}{2006}).

\bibitem[{\citenamefont{Engelbrecht}(2011)\citenamefont{Engelbrecht}}]{Engelbrecht11}
\bibinfo{author}{\bibfnamefont{A.}~\bibnamefont{Engelbrecht}},
  \bibinfo{author}{\bibfnamefont{R.} \bibnamefont{Meneses}},
  \bibnamefont{and} \bibinfo{author}{\bibfnamefont{H. J.} \bibnamefont{Sch\"ope}},
  \bibinfo{journal}{Soft Matter} \textbf{\bibinfo{volume}{7}},
  \bibinfo{pages}{5685} (\bibinfo{year}{2011}).

\bibitem{SessomsPTRSA2009} D. A. Sessoms, I. Bischofberger, L. Cipelletti and V. Trappe, Philos. Trans. R. Soc. A-Math. Phys. Eng. Sci.,  \textbf{367}, 5013, 2009.

\bibitem[{\citenamefont{Merlin}(2012)\citenamefont{Merlin}}]{Merlin12}
\bibinfo{author}{\bibfnamefont{A.}~\bibnamefont{Merlin}},
  \bibinfo{author}{\bibfnamefont{J.-B.} \bibnamefont{Salmon}},
  \bibnamefont{and} \bibinfo{author}{\bibfnamefont{J.} \bibnamefont{Leng}},
  \bibinfo{journal}{Soft Matter} \textbf{\bibinfo{volume}{8}},
  \bibinfo{pages}{3526} (\bibinfo{year}{2012}).


\bibitem[{\citenamefont{Xiong}(2006)\citenamefont{Xiong}}]{Xiong06}
\bibinfo{author}{\bibfnamefont{X. Y.}~\bibnamefont{Xiong}},
\bibinfo{author}{\bibfnamefont{K. C.}~\bibnamefont{Tam}},
  \bibnamefont{and} \bibinfo{author}{\bibfnamefont{L. H.} \bibnamefont{Gan}},
  \bibinfo{journal}{J Nanosci. Nanotechnol.} \textbf{\bibinfo{volume}{6}},
  \bibinfo{pages}{2638} (\bibinfo{year}{2006}).

\bibitem[{\citenamefont{Batrakova}(2008)\citenamefont{Batrakova}}]{Batrakova08}
\bibinfo{author}{\bibfnamefont{A. V.}~\bibnamefont{Batrakova}},
  \bibnamefont{and} \bibinfo{author}{\bibfnamefont{A. V.} \bibnamefont{Kabanov}},
  \bibinfo{journal}{J. Control Release} \textbf{\bibinfo{volume}{130}},
  \bibinfo{pages}{98} (\bibinfo{year}{2008}).


\bibitem[{\citenamefont{Wu}(1997)\citenamefont{Wu}}]{Wu97}
\bibinfo{author}{\bibfnamefont{C.}~\bibnamefont{Wu}},
  \bibinfo{author}{\bibfnamefont{T.} \bibnamefont{Liu}},
  \bibinfo{author}{\bibfnamefont{B.} \bibnamefont{Chu}},
  \bibinfo{author}{\bibfnamefont{D. K.} \bibnamefont{V.P.} \bibnamefont{Graziano}},
  \bibinfo{journal}{Macromol.} \textbf{\bibinfo{volume}{30}},
  \bibinfo{pages}{4574} (\bibinfo{year}{1997}).

  \bibitem[{\citenamefont{Rill}(1998)\citenamefont{Rill}}]{Rill98}
\bibinfo{author}{\bibfnamefont{R. L.}~\bibnamefont{Rill}},
  \bibinfo{author}{\bibfnamefont{B. R.} \bibnamefont{Locke}},
  \bibinfo{author}{\bibfnamefont{Y.} \bibnamefont{Liu}},
  \bibnamefont{and} \bibinfo{author}{\bibfnamefont{D. H.} \bibnamefont{Van Winkle}},
  \bibinfo{journal}{Proc. Natl. Acad. Sci. USA} \textbf{\bibinfo{volume}{95}},
  \bibinfo{pages}{1534} (\bibinfo{year}{1998}).

\bibitem[{\citenamefont{Svingen}(2002)\citenamefont{Svingen}}]{Svingen02}
\bibinfo{author}{\bibfnamefont{R. }~\bibnamefont{Svingen}},
  \bibinfo{author}{\bibfnamefont{R.} \bibnamefont{Alexandridis}},
  \bibnamefont{and} \bibinfo{author}{\bibfnamefont{B.} \bibnamefont{{\AA}kerman}},
  \bibinfo{journal}{Langmuir} \textbf{\bibinfo{volume}{18}},
  \bibinfo{pages}{8616} (\bibinfo{year}{2002}).

\bibitem[{\citenamefont{Svingen}(2004)\citenamefont{Svingen}}]{Svingen04}
\bibinfo{author}{\bibfnamefont{R. }~\bibnamefont{Svingen}},
  \bibnamefont{and} \bibinfo{author}{\bibfnamefont{B.} \bibnamefont{{\AA}kerman}},
  \bibinfo{journal}{J. Phys. Chem. B} \textbf{\bibinfo{volume}{108}},
  \bibinfo{pages}{2735} (\bibinfo{year}{2004}).

\bibitem[{\citenamefont{Alexandridis}(1999)\citenamefont{Alexandridis}}]{Alexandridis99}
\bibinfo{author}{\bibfnamefont{P.}~\bibnamefont{Alexandridis}},
  \bibnamefont{and} \bibinfo{author}{\bibfnamefont{R. J.} \bibnamefont{Spontak}},
  \bibinfo{journal}{Curr. Opinion Colloid Interface Sci.} \textbf{\bibinfo{volume}{4}},
  \bibinfo{pages}{130} (\bibinfo{year}{1999}).

\bibitem[{\citenamefont{Mortensen}(2001)\citenamefont{Mortensen}}]{Mortensen01}
\bibinfo{author}{\bibfnamefont{K.}~\bibnamefont{Mortensen}},
  \bibinfo{journal}{Polym. Adv. Technol.} \textbf{\bibinfo{volume}{12}},
  \bibinfo{pages}{2} (\bibinfo{year}{2001}).

\bibitem[{\citenamefont{Pozzo}(2005)\citenamefont{Pozzo}}]{Pozzo05}
\bibinfo{author}{\bibfnamefont{D. C.}~\bibnamefont{Pozzo}},
\bibinfo{author}{\bibfnamefont{K. R.}~\bibnamefont{Hollabaugh}},
  \bibnamefont{and} \bibinfo{author}{\bibfnamefont{L. M.} \bibnamefont{Walker}},
  \bibinfo{journal}{J. Rheol.} \textbf{\bibinfo{volume}{49}},
  \bibinfo{pages}{759} (\bibinfo{year}{2005}).

\bibitem[{\citenamefont{Pozzo}(2007)\citenamefont{Pozzo}}]{Pozzo07}
\bibinfo{author}{\bibfnamefont{D. C.}~\bibnamefont{Pozzo}},
  \bibnamefont{and} \bibinfo{author}{\bibfnamefont{L. M.} \bibnamefont{Walker}},
  \bibinfo{journal}{Colloids and Surfaces A: Physicochem. Eng. Aspects} \textbf{\bibinfo{volume}{294}},
  \bibinfo{pages}{117} (\bibinfo{year}{2007}).

\bibitem[{\citenamefont{Molino}(1998)\citenamefont{Molino}}]{Molino98}
\bibinfo{author}{\bibfnamefont{F. R.}~\bibnamefont{Molino}},
  \bibinfo{author}{\bibfnamefont{J.-F.} \bibnamefont{Berret}},
  \bibinfo{author}{\bibfnamefont{G.} \bibnamefont{Porte}},
  \bibinfo{author}{\bibfnamefont{O.} \bibnamefont{Diat}},
  \bibnamefont{and} \bibinfo{author}{\bibfnamefont{P.} \bibnamefont{Lindner}},
  \bibinfo{journal}{Eur. Phys. J. B} \textbf{\bibinfo{volume}{3}},
  \bibinfo{pages}{59} (\bibinfo{year}{1998}).

  \bibitem[{\citenamefont{Eiser}(2000)\citenamefont{Eiser}}]{Eiser00}
\bibinfo{author}{\bibfnamefont{E.}~\bibnamefont{Eiser}},
  \bibinfo{author}{\bibfnamefont{F.} \bibnamefont{Molino}},
  \bibnamefont{and} \bibinfo{author}{\bibfnamefont{G.} \bibnamefont{Porte}},
  \bibinfo{journal}{Eur. Phys. J. E} \textbf{\bibinfo{volume}{2}},
  \bibinfo{pages}{39} (\bibinfo{year}{2000}).


\bibitem[{\citenamefont{Ghofraniha}(2012)\citenamefont{Ghofraniha}}]{Ghofraniha12}
\bibinfo{author}{\bibfnamefont{N.}~\bibnamefont{Ghofraniha}},
  \bibinfo{author}{\bibfnamefont{E.} \bibnamefont{Tamborini}},
  \bibinfo{author}{\bibfnamefont{J.} \bibnamefont{Oberdisse}},
  \bibinfo{author}{\bibfnamefont{L.} \bibnamefont{Cipelletti}},
  \bibnamefont{and} \bibinfo{author}{\bibfnamefont{L.} \bibnamefont{Ramos}},
  \bibinfo{journal}{Soft Matter} \textbf{\bibinfo{volume}{8}},
  \bibinfo{pages}{6214} (\bibinfo{year}{2012}).


\bibitem[{\citenamefont{Meznarich}(2011)\citenamefont{Meznarich}}]{Meznarich11}
\bibinfo{author}{\bibfnamefont{N. A. K.}~\bibnamefont{Meznarich}},
   \bibnamefont{and} \bibinfo{author}{\bibfnamefont{B. J.} \bibnamefont{Love}},
  \bibinfo{journal}{Macromolecules} \textbf{\bibinfo{volume}{44}},
  \bibinfo{pages}{3548} (\bibinfo{year}{2011}).

\bibitem[{\citenamefont{Auer}(2005)\citenamefont{Auer}}]{Auer05}
\bibinfo{author}{\bibfnamefont{S.}~\bibnamefont{Auer}},
  \bibnamefont{and} \bibinfo{author}{\bibfnamefont{D.} \bibnamefont{Frenkel}},
  \bibinfo{journal}{Adv. Polym. Sci.} \textbf{\bibinfo{volume}{173}},
  \bibinfo{pages}{149} (\bibinfo{year}{2005}).

\bibitem[{\citenamefont{Alexandridis}(1995)\citenamefont{Alexandridis}}]{Alexandridis95}
\bibinfo{author}{\bibfnamefont{R.}~\bibnamefont{Alexandridis}},
  \bibinfo{author}{\bibfnamefont{T.} \bibnamefont{Nivaggioli}},
  \bibnamefont{and} \bibinfo{author}{\bibfnamefont{T. A.} \bibnamefont{Hatton}},
  \bibinfo{journal}{Langmuir} \textbf{\bibinfo{volume}{11}},
  \bibinfo{pages}{1468} (\bibinfo{year}{1995}).

\bibitem[{\citenamefont{Mortensen}(1992)\citenamefont{Mortensen}}]{MortensenPRL92}
\bibinfo{author}{\bibfnamefont{K.}~\bibnamefont{Mortensen}},
  \bibinfo{author}{\bibfnamefont{W.} \bibnamefont{Brown}},
   \bibnamefont{and} \bibinfo{author}{\bibfnamefont{B.} \bibnamefont{Nord\`{e}n}},
  \bibinfo{journal}{Phys. Rev. Lett.} \textbf{\bibinfo{volume}{68}},
  \bibinfo{pages}{2340} (\bibinfo{year}{1992}).

\bibitem[{\citenamefont{Tamborini}(2012)\citenamefont{Tamborini}}]{Tamborini12}
\bibinfo{author}{\bibfnamefont{E.}~\bibnamefont{Tamborini}},
  \bibinfo{author}{\bibfnamefont{N.} \bibnamefont{Ghofraniha}},
  \bibinfo{author}{\bibfnamefont{J.} \bibnamefont{Oberdisse}},
  \bibinfo{author}{\bibfnamefont{L.} \bibnamefont{Cipelletti}},
  \bibnamefont{and} \bibinfo{author}{\bibfnamefont{L.} \bibnamefont{Ramos}},
  \bibinfo{journal}{Langmuir} \textbf{\bibinfo{volume}{28}},
  \bibinfo{pages}{8562} (\bibinfo{year}{2012}).

\bibitem[{\citenamefont{Trong}(2008)\citenamefont{Trong}}]{Trong08}
\bibinfo{author}{\bibfnamefont{L. C. P.}~\bibnamefont{Trong}},
  \bibinfo{author}{\bibfnamefont{M.} \bibnamefont{Djabourov}},
   \bibnamefont{and} \bibinfo{author}{\bibfnamefont{A.} \bibnamefont{Ponton}},
  \bibinfo{journal}{J. Coll. Int. Sci.} \textbf{\bibinfo{volume}{328}},
  \bibinfo{pages}{278} (\bibinfo{year}{2008}).


\bibitem[{\citenamefont{Cheng}(2002)\citenamefont{Cheng}}]{Cheng02}
\bibinfo{author}{\bibfnamefont{Z.}~\bibnamefont{Cheng}},
  \bibinfo{author}{\bibfnamefont{J.} \bibnamefont{Zhu}},
  \bibinfo{author}{\bibfnamefont{P. M.} \bibnamefont{Chaikin}},
  \bibinfo{author}{\bibfnamefont{S.-E.} \bibnamefont{Phan}},
   \bibnamefont{and} \bibinfo{author}{\bibfnamefont{W. B.} \bibnamefont{Russel}},
  \bibinfo{journal}{Phys. Rev. E} \textbf{\bibinfo{volume}{65}},
  \bibinfo{pages}{041405} (\bibinfo{year}{2002}).
  
\bibitem[{\citenamefont{Segre}(1995)\citenamefont{Segr\'{e}}}]{Segre95}
\bibinfo{author}{\bibfnamefont{P. N.}~\bibnamefont{Segr\'{e}}},
  \bibinfo{author}{\bibfnamefont{S. P.} \bibnamefont{Meeker}},
    \bibinfo{author}{\bibfnamefont{P. N.} \bibnamefont{Pusey}},
   \bibnamefont{and} \bibinfo{author}{\bibfnamefont{W. C. K.} \bibnamefont{Poon}},
  \bibinfo{journal}{Phys. Rev. Lett.} \textbf{\bibinfo{volume}{75}},
  \bibinfo{pages}{958} (\bibinfo{year}{1995}).

\bibitem[{\citenamefont{Mortensen}(1992)\citenamefont{Mortensen}}]{MortensenEPL92}
\bibinfo{author}{\bibfnamefont{K.}~\bibnamefont{Mortensen}},
  \bibinfo{journal}{Europhysics. Lett.} \textbf{\bibinfo{volume}{19}},
  \bibinfo{pages}{599} (\bibinfo{year}{1992}).

\bibitem[{\citenamefont{Merlet}(2010)\citenamefont{Merlet}}]{Merlet10}
\bibinfo{author}{\bibfnamefont{N.}~\bibnamefont{Merlet-Lacroix}},
  \bibinfo{author}{\bibfnamefont{E.} \bibnamefont{Di Cola}},
   \bibnamefont{and} \bibinfo{author}{\bibfnamefont{M.} \bibnamefont{Cloitre}},
  \bibinfo{journal}{Soft Matter} \textbf{\bibinfo{volume}{6}},
  \bibinfo{pages}{984} (\bibinfo{year}{2010}).
  
  
  \bibitem[{\citenamefont{Lau}(2004)\citenamefont{Lau}}]{Lau04}
\bibinfo{author}{\bibfnamefont{B. K.}~\bibnamefont{Lau}},
  \bibinfo{author}{\bibfnamefont{Q.} \bibnamefont{Wang}},
   \bibinfo{author}{\bibfnamefont{W.} \bibnamefont{Sun}},
   \bibnamefont{and} \bibinfo{author}{\bibfnamefont{L.} \bibnamefont{Li}},
  \bibinfo{journal}{J. Polym. Sci.: Part B} \textbf{\bibinfo{volume}{42}},
  \bibinfo{pages}{2014} (\bibinfo{year}{2004}).

\bibitem[{\citenamefont{Kurz}(2005)\citenamefont{Kurz}}]{bookKurz}
\bibinfo{author}{\bibfnamefont{W.}~\bibnamefont{Kurz}},
   \bibnamefont{and} \bibinfo{author}{\bibfnamefont{D. J.} \bibnamefont{Fisher}},
  \bibinfo{journal}{Fundamentals of Solidification} \textbf{\bibinfo{volume}{}},
  \bibinfo{pages}{Trans Tech Publications, London} (\bibinfo{year}{2005}).


\bibitem[{\citenamefont{Kolmogorov}(1937)}]{Kolmogorov37}
\bibinfo{author}{\bibfnamefont{A.}~\bibnamefont{Kolmogorov}},
  \bibinfo{journal}{Izv. Akad. Nauk SSSR, Ser. Fiz.} \textbf{\bibinfo{volume}{3}},
  \bibinfo{pages}{355} (\bibinfo{year}{1937}).

\bibitem[{\citenamefont{Johnson}(1939)}]{Johnson39}
\bibinfo{author}{\bibfnamefont{W. A.}~\bibnamefont{Johnson}} \bibnamefont{and}
  \bibinfo{author}{\bibfnamefont{R. F.}~\bibnamefont{Mehl}},
  \bibinfo{journal}{Trans. Am. Inst. Min., Metall. Pet. Eng.} \textbf{\bibinfo{volume}{135}},
  \bibinfo{pages}{416} (\bibinfo{year}{1939}).

\bibitem[{\citenamefont{Avrami}(1939)}]{Avrami39}
\bibinfo{author}{\bibfnamefont{M.}~\bibnamefont{Avrami}},
  \bibinfo{journal}{J. Chem. Phys.} \textbf{\bibinfo{volume}{7}},
  \bibinfo{pages}{1103} (\bibinfo{year}{1939}).

\bibitem[{\citenamefont{Avrami}(1940)}]{Avrami40}
\bibinfo{author}{\bibfnamefont{M.}~\bibnamefont{Avrami}},
  \bibinfo{journal}{J. Chem. Phys.} \textbf{\bibinfo{volume}{8}},
  \bibinfo{pages}{212} (\bibinfo{year}{1940}).

  \bibitem[{\citenamefont{Avrami}(1941)}]{Avrami41}
\bibinfo{author}{\bibfnamefont{M.}~\bibnamefont{Avrami}},
  \bibinfo{journal}{J. Chem. Phys.} \textbf{\bibinfo{volume}{9}},
  \bibinfo{pages}{177} (\bibinfo{year}{1941}).

\bibitem[{\citenamefont{Farjas}(2007)}]{Farjas07}
\bibinfo{author}{\bibfnamefont{J.}~\bibnamefont{Farjas}} \bibnamefont{and}
  \bibinfo{author}{\bibfnamefont{P.}~\bibnamefont{Roura}},
  \bibinfo{journal}{Phys. Rev. B} \textbf{\bibinfo{volume}{75}},
  \bibinfo{pages}{184112} (\bibinfo{year}{2007}).

\bibitem[{\citenamefont{Farjas}(2008)}]{Farjas08}
\bibinfo{author}{\bibfnamefont{J.}~\bibnamefont{Farjas}} \bibnamefont{and}
  \bibinfo{author}{\bibfnamefont{P.}~\bibnamefont{Roura}},
  \bibinfo{journal}{Phys. Rev. B} \textbf{\bibinfo{volume}{78}},
  \bibinfo{pages}{144101} (\bibinfo{year}{2008}).

\bibitem[{\citenamefont{Kelton}(2010)\citenamefont{Kelton}}]{bookKelton}
\bibinfo{author}{\bibfnamefont{K. F.}~\bibnamefont{Kelton}},
   \bibnamefont{and} \bibinfo{author}{\bibfnamefont{A. L.} \bibnamefont{Greer}},
  \bibinfo{journal}{Nucleation in Condensed Matter. Applications in Materials and Biology} \textbf{\bibinfo{volume}{}},
  \bibinfo{pages}{Elsevier, Amsterdam} (\bibinfo{year}{2010}).

\bibitem[{\citenamefont{Zhang}(2005)}]{Zhang05}
\bibinfo{author}{\bibfnamefont{H.}~\bibnamefont{Zhang}},
 \bibinfo{author}{\bibfnamefont{I.} \bibnamefont{Hussain}},
 \bibinfo{author}{\bibfnamefont{M.} \bibnamefont{Brust}},
 \bibinfo{author}{\bibfnamefont{M. F.} \bibnamefont{Butler}},
 \bibinfo{author}{\bibfnamefont{S. P.} \bibnamefont{Rannard}}, \bibnamefont{and}
  \bibinfo{author}{\bibfnamefont{A. I.}~\bibnamefont{Cooper}},
  \bibinfo{journal}{Nature Materials} \textbf{\bibinfo{volume}{4}},
  \bibinfo{pages}{787} (\bibinfo{year}{2005}).

\bibitem[{\citenamefont{Lee}(1999)}]{Lee99}
\bibinfo{author}{\bibfnamefont{Y. C.}~\bibnamefont{Leer}},
 \bibinfo{author}{\bibfnamefont{A. K.} \bibnamefont{Dahle}},
   \bibinfo{author}{\bibfnamefont{D. H.} \bibnamefont{StJohn}}, \bibnamefont{and}
  \bibinfo{author}{\bibfnamefont{J. E. C.}~\bibnamefont{Hutt}},
  \bibinfo{journal}{Materials Science and Engineering A} \textbf{\bibinfo{volume}{259}},
  \bibinfo{pages}{43} (\bibinfo{year}{1999}).

\bibitem[{\citenamefont{Miller}(1987)}]{Miller87}
\bibinfo{author}{\bibfnamefont{A.}~\bibnamefont{Miller}}, \bibnamefont{and}
  \bibinfo{author}{\bibfnamefont{H.}~\bibnamefont{M\"ohwald}},
  \bibinfo{journal}{J. Chem. Phys.} \textbf{\bibinfo{volume}{86}},
  \bibinfo{pages}{4258} (\bibinfo{year}{1987}).


\bibitem{BrambillaPRL2009} G. Brambilla, D. El Masri, M. Pierno, L. Berthier, L. Cipelletti, G. Petekidis and A. B. Schofield, Phys. Rev. Lett.,  \textbf{102}, 085703, 2009.



\bibitem[{\citenamefont{HernandezGuzman}(2009)}]{HernandezGuzman09}
\bibinfo{author}{\bibfnamefont{J.}~\bibnamefont{Hern\'andez-Guzm\'an}}, \bibnamefont{and}
  \bibinfo{author}{\bibfnamefont{E. R.}~\bibnamefont{Weeks}},
  \bibinfo{journal}{Proc. Natl. Acad. Sci. USA} \textbf{\bibinfo{volume}{106}},
  \bibinfo{pages}{15198} (\bibinfo{year}{2009}).

\bibitem[{\citenamefont{Ramsteiner}(2010)\citenamefont{Ramsteiner}}]{Ramsteiner10}
\bibinfo{author}{\bibfnamefont{I. B.}~\bibnamefont{Ramsteiner}},
  \bibinfo{author}{\bibfnamefont{D. A.} \bibnamefont{Weitz}}, \bibnamefont{and}
  \bibinfo{author}{\bibfnamefont{F.} \bibnamefont{Spaepen}},
  \bibinfo{journal}{Phys. Rev. E} \textbf{\bibinfo{volume}{82}},
  \bibinfo{pages}{041603} (\bibinfo{year}{2010}).

  \bibitem[{\citenamefont{Nguyen}(2011)\citenamefont{Nguyen}}]{Nguyen11}
\bibinfo{author}{\bibfnamefont{V. D.}~\bibnamefont{Nguyen}},
  \bibinfo{author}{\bibfnamefont{Z.} \bibnamefont{Hu}}, \bibnamefont{and}
  \bibinfo{author}{\bibfnamefont{P.} \bibnamefont{Schall}},
  \bibinfo{journal}{Phys. Rev. E} \textbf{\bibinfo{volume}{84}},
  \bibinfo{pages}{011607} (\bibinfo{year}{2011}).




\end{thebibliography}
%\bibliographystyle{unsrt}

\end{document}